\documentclass[prd,11pt,superscriptaddress,onecolumn,preprintnumbers,
showpacs,showkeys,nofootinbib]{revtex4-1}
\usepackage{amssymb,amsmath,graphicx,epsfig,subfig,color,hyperref,url} 
\begin{document}
\tolerance=100000
\thispagestyle{empty}
\setcounter{page}{0}

\newcommand{\be}{\begin{equation}}
\newcommand{\ee}{\end{equation}}
\newcommand{\br}{\begin{eqnarray}}
\newcommand{\er}{\end{eqnarray}}
\newcommand{\ba}{\begin{array}}
\newcommand{\ea}{\end{array}}
\newcommand{\bi}{\begin{itemize}}
\newcommand{\ei}{\end{itemize}}
\newcommand{\bn}{\begin{enumerate}}
\newcommand{\en}{\end{enumerate}}
\newcommand{\bc}{\begin{center}}
\newcommand{\ec}{\end{center}}
\newcommand{\ul}{\underline}
\newcommand{\ra}{\rightarrow}
\newcommand{\susy}{{{SUSY}}}

\def \pt{p{\!\!\!/}_T}
\def \lumi{{\cal L}}  
\def\invfb{fb^{-1}} 
\def \MCH {$\tilde\chi_1^+$}
\def \CH{{\tilde\chi}^{\pm}}
\def \N0{\tilde\chi^0}
\def \LSP{\tilde\chi_1^0}
\def \SNU{\tilde{\nu}}
\def \BARSNU{\tilde{\bar{\nu}}}
\def \MLSP{m_{{\tilde\chi_1}^0}}
\def \MCH{m_{{\tilde\chi}^{\pm}}}
\def \MCHMIN {\MCH^{min}}
\def \ET{\not\!\!{E_T}}
\def \LL{\tilde{l}_L}
\def \LR{\tilde{l}_R}
\def \MLL{m_{\tilde{l}_L}}
\def \MLR{m_{\tilde{l}_R}}
\def \MSNU{m_{\tilde{\nu}}}
\def\Ecm{\ifmmode{E_{\mathrm{cm}}}\else{$E_{\mathrm{cm}}$}\fi}
\def\gluino{\ifmmode{\mathaccent"7E g}\else{$\mathaccent"7E g$}\fi}
\def\photino{\ifmmode{\mathaccent"7E \gamma}\else{$\mathaccent"7E \gamma$}\fi}
\def\gl{\ifmmode{m_{\mathaccent"7E g}}
             \else{$m_{\mathaccent"7E g}$}\fi}
\def\taugluino{\ifmmode{\tau_{\mathaccent"7E g}}
             \else{$\tau_{\mathaccent"7E g}$}\fi}
\def\mphotino{\ifmmode{m_{\mathaccent"7E \gamma}}
             \else{$m_{\mathaccent"7E \gamma}$}\fi}
\def\ML{\ifmmode{{\mathaccent"7E M}_L}

             \else{${\mathaccent"7E M}_L$}\fi}
\def\MR{\ifmmode{{\mathaccent"7E M}_R}
             \else{${\mathaccent"7E M}_R$}\fi}
\def\lsim{\buildrel{\scriptscriptstyle <}\over{\scriptscriptstyle\sim}}
\def\gsim{\buildrel{\scriptscriptstyle >}\over{\scriptscriptstyle\sim}}
\newcommand{\comment}[1]{}

\begin{titlepage}

\begin{center}
{\LARGE\bf
Looking for an Invisible Higgs Signal at the LHC} \\[5mm]
\bigskip
{\large\sf Diptimoy Ghosh} $^{a}$, 
{\large\sf Rohini Godbole} $^{b}$, 
{\large\sf Monoranjan Guchait} $^{c}$,
{\large\sf Kirtimaan Mohan} $^{b}$
and
{\large\sf Dipan Sengupta} $^{c}$ 
\\ [4mm]
\bigskip

{\noindent $^{a)}$ 
INFN, Sezione di Roma,\\
\hspace*{0.1in} Piazzale A.  Moro 2, I-00185 Roma, Italy. }  \\
 
\medskip

{\noindent $^{b)}$ 
Center for High Energy Physics, 
Indian Institute of Science,  \\
\hspace*{0.1in} Bangalore, 560012, India. }

\medskip

{\noindent $^{c)}$ 
Department of High Energy Physics, 
Tata Institute of Fundamental Research,  \\
\hspace*{0.1in} 1, Homi Bhabha Road, Mumbai 400 005, India.  }
\end{center}

\begin{abstract}
 {While the  recent discovery of 
 a Higgs-like boson  at the LHC is 
 an extremely important and encouraging step
 towards the   discovery of the {\it complete}
standard model(SM), the current information on this state does
 not rule out  possibility of 
beyond standard model (BSM)  physics.  
In fact the current data can still accommodate reasonably large 
values of the branching fractions of the Higgs into
a channel with `invisible' decay products, such a channel being
also well motivated theoretically. In this study we revisit the
possibility of detecting  the Higgs in this
invisible  channel for both choices of the LHC energies, 8 and
 14 TeV, for two production modes; vector boson fusion(VBF)
  and associated production($ZH$). We perform a comprehensive
 collider analysis for all the above channels and project the
 reach of LHC to constrain the invisible decay branching
 fraction for both 8 and 14 TeV energies. For the ZH case we consider decays
 of the $Z$ boson into a pair
 of leptons as well as a $b \bar b$ pair. For the VBF channel
 the sensitivity is found to be more than $5 \sigma$ for both the
 energies  up to an invisible branching ratio ($\rm { B}r_{inv})
 \sim 0.80$, with luminosities $\sim 20/30 {\rm fb}^{-1}$. The
 sensitivity is further extended to  values of $\rm { B}r_{inv}
 \sim 0.25$ for $300~{\rm fb}^{-1}$ at $14$ TeV. However the
 reach is found to be more modest for the  $ZH$ mode with 
leptonic final state; with about $3.5 \sigma$ for the planned
 luminosity at $8$ TeV, reaching $8 \sigma$ only for $14$ TeV
 for $50~ {\rm fb}^{-1}$. In spite of the much larger branching
 ratio (BR) of the $Z$  into a $b \bar b$ channel compared to
 the dilepton case, the former channel, can provide  useful
 reach up to $\rm Br_{inv} \gsim 0.75$, only for the higher 
luminosity ($300~{\rm fb}^{-1}$) option using both jet-substructure
 and jet clustering methods.}

\normalsize

\end{abstract}

\maketitle

\end{titlepage}

\section{Introduction}
 The unprecedented high precision to which the Standard 
Model (SM) \cite{Glashow:1961tr,Weinberg:1967tq,Salam:1968rm}
 has been tested as well as the discovery of a Higgs
like boson at both the ATLAS and CMS~\cite{ATLAS:2012gk,CMS:2012gu}
 notwithstanding,  the deficiencies of the Standard model (SM)
 both of the  observational~\cite{Bertone:2004pz,Altarelli:2004za} 
and aesthetic 
 justify the existence of physics beyond the Standard
Model (BSM). For example, the instability of the electroweak
scale under radiative corrections is cured by several BSM
 models.  So far the Large Hadron Collider (LHC) has not given
 us any evidence of BSM physics.  The recent discovery of a
 Higgs like boson at both the ATLAS and CMS~\cite{ATLAS:2012gk,CMS:2012gu}
 experiments has opened up new avenues for discovering or 
restricting the possibility of various BSM scenarios.  
 Since many of the extensions of the SM
have been suggested to  address the issue of stability of the 
electroweak symmetry breaking scale against radiative corrections, 
all of them have implications for properties of the Higgs sector
 such  as the  number of the Higgs bosons and their couplings and CP 
properties.   Hence it is very likely that a 
study of the properties of this boson can also yield information 
about BSM physics.  A clear understanding of the characteristics
 of the Higgs will elucidate not only the nature of electroweak
 symmetry breaking (EWSB), but also help in our understanding of
 how a BSM spectrum may generate or be part of EWSB.  
The lack of appearance at the LHC of any other particle, not expected 
in the SM, so far in fact means that  the  properties
of the Higgs may therefore, give us our first glimpse at BSM physics. 
 
Of course the first important step in establishing the new boson which 
we have discovered  as `a' Higgs boson, will be to have some 
pointers to the spin and CP of the state. 
However,  finally, the identification of this  boson
as the particle responsible for EWSB requires the determination of
its coupling to fermions and gauge bosons. Let us note that the tree 
level couplings of the Higgs  to the fermions and the electroweak 
gauge bosons, are completely determined by 
the details of the EWSB. On the other hand the loop induced 
couplings of the Higgs to a pair of $gg$ and $\gamma \gamma$,
 as well as the higher dimensional operators in other couplings
 can receive contributions from BSM physics as well. Hence,
 the measurement of the relative decay widths of the Higgs into
 different final states will not only provide information  about
 the EWSB mechanism but  may also carry with it information
about BSM particle spectra.  For example, the apparent excess of
events seen in the  $H\rightarrow \gamma \gamma$ channel coupled
with  non observation in the $H \rightarrow \tau \tau$ 
channel~\cite{ATLAS:2012gk,CMS:2012gu}, if confirmed, 
will have strong implications for various BSM models. 

Strong cosmological evidence supporting the existence of 
Dark Matter (DM)  means that  almost all extensions of SM
must include in their spectra a candidate for it which is supposed 
to be neutral 
and weakly interacting. A large number
of such models allow for a significant branching fraction for 
the decay of the Higgs to DM, thus providing a channel where the
Higgs decay is ``invisible'' to the detector.
 In the SM the Higgs can decay invisibly  through 
$H\rightarrow ZZ^*\rightarrow4\nu$, which can 
only contribute to roughly $0. 1\%$ of the branching 
ratio~\cite{Denner:2011mq}.  Therefore, the observation of a
 sizable invisible branching ratio ($\rm Br_{inv}$) of
 the Higgs will be a strong indication for BSM physics.
 There exist several examples of BSM physics models where the Higgs can
 have an invisible decay, such as, the decay of the Higgs to
 the lightest supersymmetric particle (LSP)~\cite
{Belanger:2001am}, decay to graviscalars in extra-dimensional
 models~\cite{Giudice:2000av,Battaglia:2004js} in gauge 
extensions of the SM~\cite{Gopalakrishna:2009yz,Drozd:2011aa}
and in models for neutrino masses~\cite{Ghosh:2011qc,Belotsky:2002ym,Shrock:1982kd}.
It has been noticed in various analysis
\cite{Espinosa:2012vu,Carmi:2012in,Giardino:2012dp,Dobrescu:2012td}
that if this resonance is interpreted as a Higgs boson, the currently
available information on its properties can allow non trivial 
 values of $\rm Br_{inv}$.  
In fact a recent analysis by the CMS collaboration performing  
a global fit  to the LHC data, suggests that an invisible 
branching ratio in non SM channel as large as $62 \%$ at 95$\%$
 confidence level of the Higgs of mass $\sim$125 GeV
is still allowed~\cite{CMS-PAS-HIG-12-045}
\footnote{Similarly ATLAS also obtained a lower bound of 84$\%$
 at 95 $\%$ confidence level on
the invisible branching ratio of the Higgs
without any assumption on the total  decay width
\cite{ATLAS-CONF-2012-127}.}. In fact detailed analysis of
 LEP data showed no evidence for an invisibly decaying Higgs of
 mass less than 112.1~GeV \cite{Searches:2001ab}.

The feasibility of 
determining an invisible branching fraction of the Higgs for 
$\sqrt{s}=$ 7 TeV,8 TeV and 14 TeV at the
 LHC has been studied in various production modes of the  Higgs 
~\cite{Gunion:1993jf,Choudhury:1993hv,Eboli:2000ze,Godbole:2003it,
Barger:2012hv,DiGirolamo:2001yv,Davoudiasl:2004aj,Englert:2011us,
Djouadi:2012zc,Bansal:2010zz,Bai:2011wz}
 which is described very briefly in the next section.  
We look at the production of Higgs in association with a 
electroweak gauge boson as well as through Vector Boson 
Fusion (VBF) in detail. In earlier studies, the leptonic decay
 of the $ Z $ boson was used to identify the invisible decay of a 
Higgs produced in association with a $Z$ 
boson~\cite{Godbole:2003it}.  In the present study we update the 
analysis in the leptonic channel and also probe the 
possibility of detecting an invisible decay of the Higgs
 by identifying the associated $Z$ boson through b-tagged jets
 both for 8 TeV and as well as 14 TeV LHC.
 We also apply the jet-substructure algorithm~
\cite{Butterworth:2008iy} for b-tagged  final states which 
marginally help in improving signal acceptance efficiencies.  
In addition, we study how the invisible decay channel can be
 probed in the production of the Higgs via vector boson fusion
 for both 8 TeV and 14 TeV LHC energy. 

We organize our work as follows.
In section 2, we discuss very briefly about the invisible decay 
of Higgs.  In the subsequent sections 3 and 4, we describe 
simulation of invisible Higgs signal for VBF  and ZH channels. 
Finally, we summarize our observations in section 5.
  
\section{Signatures of an invisibly decaying Higgs}
\begin{figure} 
 \centering
 \scalebox{0.55}{\includegraphics{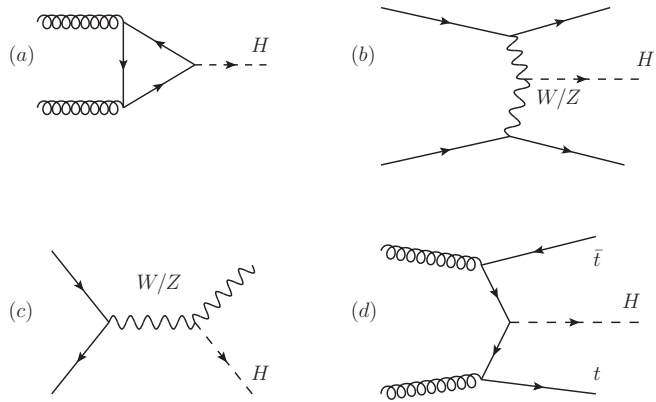}}
 \caption{\label{fdiag} Higgs production channels at the LHC:
(a)gluon-gluon fusion(ggF), (b) Vector boson fusion(VBF), 
associated productions (c) ZH and (d) $t\bar t H$.}
 \end{figure}

There are four main production mechanisms of the Higgs boson in a
 hadron collider. The most  dominant one is gluon-gluon fusion
 via a top quark loop (ggF)  $(gg \rightarrow H )$ 
followed by  VBF
$(q \bar q \rightarrow q\bar q H)$, then
 Higgs production in
 association with vector bosons (VH)  
$(q \bar q \rightarrow ZH / WH)$ and finally in association
 with top quark pairs (ttH) $(gg/q\bar q \rightarrow t\bar t H)$ with
the lowest cross section
~\cite{PhysRevLett.40.692,PhysRevLett.70.1372,Spira:1995rr,Harlander:2005rq,
Dittmaier:2011ti,Dawson:1990zj, 
Djouadi:1991tka,Harlander:2002wh,Anastasiou:2002yz,Ravindran:2003um,Cahn:1983ip,
r1,r2,
Hikasa1985385,Altarelli1987205,PhysRevLett.69.3274,Figy:2003nv,Ciccolini:2007jr,
Bolzoni:2010xr,Harlander:2008xn,
Figy:2004pt,Berger:2004pca,PhysRevD.18.1724,Kunszt1991247,
Han1991167,Hamberg:1990np,Ciccolini:2003jy,PhysRevD.47.2730,PhysRevD.47.2722,
Raitio:1978pt,PhysRevD.29.876,Kunszt1984339,Gunion1991510,PhysRevLett.66.2433,Dittmaier:2003ej,
Dawson:2003kb,Reina:2001sf,
Beenakker:2001rj,Dawson:2003zu,Baglio:2010ae}. 
The various production channels are shown in
Fig.~\ref{fdiag}.  
Needless to say that the signatures of the Higgs particle are 
characterized  by the pattern of the Higgs decay 
channels \cite{Djouadi:2005gi}. 
  Recall that the BR
 of the Higgs decay in the invisible channel in the framework of SM
 is too low to be observed, therefore, any observation of invisible
 decay channel of the Higgs will shed some light about BSM physics. 
On the other hand the production cross section of the Higgs can
 vary in various models due to the presence of new particles 
inside loops and  modified couplings of Higgs with gauge bosons
 and fermions.  For example,  supersymmteric (SUSY) particles
 may alter the loop contribution in ggF channel~
\cite{Carena:1995bx,Dawson:1996xz,Harlander:2003bb,Harlander:2003kf,Harlander:2004tp,Harlander:2005if,
Degrassi:2008zj,Degrassi:2011vq,Muhlleitner:2008yw,Degrassi:2010eu,Harlander:2010wr,Anastasiou:2006hc,
Aglietti:2006tp,Bonciani:2007ex,Muhlleitner:2006wx,Anastasiou:2008rm,Pak:2010cu,
Djouadi:2005gj}. 
Consequently signal in the invisible decay channel will be a combined 
effect due to the modified Higgs production cross section and its
 branching ratio in the invisible channel. Hence 
 this makes it difficult to constrain only the invisible decay
 branching ratio of the  Higgs $\rm BR_{inv}(H\rightarrow inv)$. Instead what
 can be constrained is in fact 
\begin{equation}
 R_{inv}\equiv\sigma_H^{BSM}BR(H\rightarrow inv)/\sigma_H^{SM}
\label{inveq}
\end{equation}
where $\sigma_H^{BSM}$ and $\sigma_H^{SM}$ stand for the Higgs
 production cross sections in the framework of corresponding BSM
 and SM respectively. At leading order, the Higgs produced through
 ggF and decaying invisibly would be hard to detect because of
 soft  missing transverse momentum ($p{\!\!\!/}_T$).  
However, at higher orders in QCD  for ggF, the Higgs can be
 produced in association with a single jet and one can 
then look for a considerably large missing transverse momentum
along with a jet. Interestingly, such final states with a 
mono-jet have been analyzed with $1~ \rm fb^{-1}$ of data at 
$\sqrt{s}=7 $~TeV for both CMS~\cite{cmsmonojet} and
 ATLAS~\cite{Aad:2011xw}. Using those results, 
 $R_{inv}$ in eq. \ref{inveq} can be constrained and is
 found to be more than 10 at 95$\%$ CL with $1 \rm fb^{-1}$
 data~\cite{Englert:2011us}. Moreover, the mono-jet 
search has also been analyzed by including a second hard 
jet~\cite{cmsmonojet} thus also including events from VBF and
 VH processes in the signal.  It has been argued recently that
 at $4. 7~ \rm fb^{-1}$ data 
at $\sqrt{s}=7 $~TeV, this can be reduced to $R_{inv}<2$ and 
for $15 ~\rm fb^{-1}$ of data at $8$~ TeV this can be further 
reduced to $R_{inv}<0. 9$ \cite{Djouadi:2012zc}.
One should note here that even though the production 
cross-section is large the mono-jet searches are plagued by large
 $V$+jets ($V=W,Z$) background (Bg). 

The most promising channel for the detection of an invisibly decaying 
Higgs is VBF since it has a relatively large cross section and has
 an  unique event topology that can be used to effectively 
remove backgrounds~\cite{Eboli:2000ze,DiGirolamo:2001yv,Bansal:2010zz}. 
 The signal consists of jets moving in opposite directions with large 
rapidity gaps.  A recent study has shown that $R_{inv}$ as low as $0. 21$ 
can be probed with $30~\rm fb^{-1}$ data at $\sqrt{s}=14$~TeV and for 
$\sqrt{s}=7$~TeV with $20~\rm fb^{-1}$ it can be 
probed to as low as $0. 4$ with $95 \%$ CL~\cite{Bai:2011wz}. 
 
We revisit this analysis for 8 TeV and 14 TeV energies. 
In our current analysis we employ a different set  
of kinematic selection cut values to that used in Ref~\cite{Bai:2011wz}.
Moreover, in this analysis a precise method of jet reconstruction 
with $\rm anti-k_{T}$ \cite{Cacciari:2008gp} algorithm 
built in the FastJet \cite{Cacciari:2011ma} package is implemented.
It has to be noted that in our analysis we consider the additional
W/Z+3 jets backgrounds which were not considered in 
 earlier works \cite{Bai:2011wz,Cavalli:2002vs}. These additional modes do
 contribute a sizable fraction to the total 
background cross section, in particular Z+3 jets channel. As a consequence, 
 our conclusion appears to be
different than previous works \cite{Bai:2011wz,Cavalli:2002vs} which is
discussed in  Sec.3.
However, the main drawback of VBF channels is that it has large 
systematic uncertainties and it is difficult to estimate the QCD 
background \cite{Bansal:2010zz,DiGirolamo:2001yv}. 

The $t\bar{t}H$ channel has been studied in detail~\cite{Kersevan:2002zj} 
for $\sqrt{s}=14$~TeV LHC in both the semileptonic, 
$t \bar t \rightarrow Wb Wb \rightarrow l \nu b q \bar q b$, and as well 
as in the hadronic mode 
$\rightarrow q \bar q b q \bar q b$.  The complex final state and the 
combinatorial background requires a very sophisticated 
analysis. 

The cleanest channel by far is the associated production channel
$VH(V=W,Z)$.  
Incidentally, the couplings between gauge boson and Higgs are not
expected to deviate from the SM significantly because of the 
unitarity of  the theory and restrictions from electroweak precision tests. 
As a consequence, in any BSM model, the parton level cross sections 
for VBF and ZH channels turn out to be very close to SM values.  
 Under the assumption that the Higgs gauge couplings do not deviate 
from  standard model couplings, 
 these channels therefore
give a direct probe of the invisible branching ratios, unlike
ggF.
  However, the $WH$ channel is diluted by the inclusive $W$
background which makes it difficult to use for detecting an invisible Higgs 
decay~\cite{Gagnon:2009zz} where as the $ZH$ channel is more 
promising because  of the presence of two leptons from the  $Z$  boson decay. 
 We study here the efficacy of this channel in detecting invisible
 branching  ratio at $\sqrt{s}=8$~TeV and  
$\sqrt{s}=14$~TeV energies.   Like earlier studies of this 
channel~\cite{Godbole:2003it,Bai:2011wz,Gagnon:2009zz}  for 14 TeV
energy, we use the leptonic decay to identify the $Z$ boson. 
In addition to revisiting this channel for 14 TeV energy, we analyze
it for 8 TeV energy which are the new results for this channel.
Moreover, we consider
the hadronic decay mode, specifically decay to $b$ quarks and investigate 
the viability of use of jet substructure and clustering methods for 
detection of $b$ jets in reducing backgrounds.

\section{Invisible Higgs signal via VBF}
In this section we study the feasibility of 
finding the invisible Higgs signal through 
the VBF process which is the sub dominant process
for the Higgs production in hadron colliders.  
This channel has been studied previously for 14 TeV LHC 
\cite{Eboli:2000ze,DiGirolamo:2001yv,Djouadi:2012zc,Gagnon:2009zz}
 and very recently for 7 TeV and 8 TeV \cite{Bai:2011wz}.  
 We also revisit this analysis for 8 TeV
 and 14 TeV LHC energy for the  Higgs mass of 125 GeV  
using a different set of  selection cut values.  
In this channel, the Higgs  is produced through vector boson
 fusion, where vector bosons originate by radiating off
 two initial quarks along with two jets,

\br
{\bf  p p \to q q h \to
 2 jets +p{\!\!\!/}_T}.
\er

The final state consists of two jets in the forward and backward
 directions with a wide separation in rapidity and a reasonably
 large $\pt$ due to the presence of non-interacting particles
 from Higgs decay. In addition to this pure VBF processes, there
 are some non VBF processes which also provide the same final 
state consisting of 2 jets and $\pt$. For instance, higher order
 QCD effects in ggF process can give rise to two jets in the final
 states because of a hard emission of partons from the initial
 states with  a non negligible cross section.   
The dominant SM background processes for this signal are due to 
($ W \to \ell \nu$)+jets, ($Z \to \nu\bar{\nu}$)+jets,
$t\bar{t}$ ($tbW$) and QCD.  For $W$+jets,
a significant background can arise if the lepton is not detected. 
 Note that the background cross sections mimicking the signal
are significantly large, and hence a sizable reduction is 
required to achieve a reasonable signal sensitivity. 
The  signal and background processes are simulated using
{\tt MadGraph/Madevent} \cite{Alwall:2011uj} and 
subsequently passed through {\tt PYTHIA6} \cite{Sjostrand:2006za} for 
parton showering. In this study for all numerical calculations  
we use CTEQ6L~\cite{Pumplin:2002vw} for parton distribution functions.  
In the process of showering, we adopt
MLM matching\cite{Hoche:2006ph} using default
values set by the {\tt MadGraph/MadEvent} suite 
to avoid double counting of jets. 
Jets are reconstructed using {\tt FastJet} \cite{Cacciari:2011ma}
with anti-$\rm K_{T}$ \cite{Cacciari:2008gp} algorithm using size
 parameter $R =0. 5$ and applying a jet $ p_{T}$ threshold of
 40 GeV and $|\eta|<4.5$.  Notice that the signal is completely
 free from leptonic activities whereas background channels may
 contain leptons in the final state.  Therefore, a leptonic veto
 might help to eliminate certain fraction of backgrounds. 
Leptons are selected with $p_T^\ell >$10~GeV, $|\eta_\ell|<$2.5. 
We compute missing transverse energy from the momenta of all
visible particles.  The following set of cuts are used in the 
simulation :
\begin{enumerate}
\item 
VBF selections: 
The leading jets in Higgs production through the
VBF process are produced in the forward and backward direction 
and hence is expected to have a large rapidity gap. 
Therefore, we select events where the absolute rapidity
difference between the two leading jets is  
$|\eta_{j1} - \eta_{j2}|=|\Delta \eta|>$4. 
To ensure that the two jets are in  the opposite direction,
 the product of rapidity of two jets are required to be, 
$\eta_{j1} \times \eta_{j2} <$0. 

\item
Central Jet veto: For a pure VBF process, no jets with $p_{T}>40$ GeV
 are expected in the rapidity gap region between two reconstructed jets.
 Therefore we discard events if there
 be any jets in  central region.

\item
Lepton veto(LV): Since the signal has a pure hadronic topology,
 events with any lepton are vetoed out. 

\item
Selection of $p{\!\!\!/}_T$: Events are required to have at least
 $\pt>$100 (170)~GeV for 8 (14) TeV energy. 
\item
Dijet invariant mass $M_{jj}$ :
The invariant mass of two leading jets is expected to be very
 large and hence we demand, $M_{jj}>$1400 (1800)~GeV for 8 
TeV (14 TeV) energy.   
\end{enumerate} 
\begin{table}[t]
\begin{center}
\begin{tabular}{|l|lr|lr| }
\hline
Process & \multicolumn{2}{c|}{8 TeV} & \multicolumn{2}{c|}{14 TeV} \\
  & Production  & After cuts & Production & After cuts\\
  & CS[pb]  &  CS[fb]  & CS[pb] & CS[fb]   \\    
\hline
W+2jets(VBF) & 76.5  & 4.5    & 167.9 & 6.3 \\
W+2jets & 18700    &  5.8 & 45900  & 18.7\\
W+3jets & 10260       &   $<1$ & 21000  & 13 \\
Z+2jets(VBF) & 19 & 6 & 43.2  &6.7 \\
Z+2jets & 6000 & 16.5  & 14000 & 11.2 \\
Z+3jets & 2772 & 8.3 & 7300 & 17.8   \\
tbW & 140 & $<1$ & 611  & $<1$ \\
\hline
Total Background & & 41.1 &  & 74\\
\hline
hjj(VBF) & 1.73 & 7.3  & 4.3 & 8.7\\
hjj & 6.7 & 1.2 & 24.5 & 1.3\\
\hline
Signal & & 8.5 & & 10\\
\hline
\end{tabular} 
\caption{ 
Event summary of the signal and backgrounds for the final state with
two jets and $\pt$ via VBF channel for 8 TeV and 14 TeV LHC energies.  
In the second column the cross sections corresponding to 
production and after all cuts are shown for signal and background 
processes respectively for 8 TeV energy. 
The third column presents the same for 14 TeV energy.}
\label{vbf} 
\end{center}
\end{table} 
We notice that $\pt$ and $M_{jj}$ cuts are extremely useful to suppress the 
backgrounds with a marginal effect in the signal cross section.
We have also checked that the background contribution due to QCD is
 negligible because of a strong $\pt$ and a large di-jet invariant mass 
cut($M_{jj}$); this is why results for QCD are not presented here.
 In our simulation, the rejection efficiencies due to the
central jet veto for QCD Wjj and QCD Zjj are about $~20\%$
for both energies.
Note that this efficiency depends crucially 
on the detector effects like calibrations, electronic noise,
 pile up effects etc. \cite{Cavalli:2002vs},
which are not taken into account in this analysis.  

In Table \ref{vbf} we present the event summary for signal and 
all background processes subjected to the above set of cuts.  
The first column represents the production  cross section at the
leading order obtained from {\tt MadGraph} \cite{Alwall:2011uj}.  
The contribution due to the pure  VBF type and non-VBF type
 of processes are shown separately.  In the subsequent columns,
 the  cross sections  subject to all cuts are presented.  Notably,
 there exists a non negligible possibility that $ W/Z$+3jet 
channel may contribute to the background cross section, if
 the third jet is not detected.  Here we  present our results for both the  
8 TeV and 14 TeV energies. At 8 TeV energy, the total signal cross section
turns out to be 8.5 fb,
consisting of 14 $\%$ contribution from ggF and the rest due to VBF process. 
 At 8 TeV energy, for 
$\lumi$=20$~\rm \invfb$, it is possible to observe signal
with S/$\sqrt{B}\sim$5.9  leading to a detection of invisible 
BR $\sim$84\% or above assuming $\sigma_{SM}$ = $\sigma_{BSM}$
in Eq.~\ref{inveq}.  On the other hand, for 14 TeV energy, results
 are more encouraging where one can find a signal with a
better sensitivity yielding S/$\sqrt{B} \sim$ 6.3 (20) for 30 (300)
$\rm \invfb$ integrated luminosity which predicts a measurement of BR
 $\ge$ 0.79(0.25).  In our estimation the signal purity S/(S+B) is
 approximately 40\% 
lower than the results obtained by the Ref.\cite{Bai:2011wz}.
 As mentioned earlier, we use a more reliable method of jet reconstruction
by using FastJet \cite{Cacciari:2011ma} with $\rm anti-K_T$ algorithm
 \cite{Cacciari:2008gp},
and consider an additional
 W/Z+3jets background.  
It is to be noted that in our calculation we
used LO cross sections for both signal and backgrounds. However the
K-factor for the signal is $\sim$0.95 \cite{Figy:2010ct} and for
$W/Z$+jets it is also very
close to 1($\sim$1.1) \cite{Berger:2010zx,Ita:2011wn}.  Therefore, 
inclusion of K-factors  in the above calculation will not
alter the conclusions significantly.

\section{Invisible Higgs signal via $ZH$}
Here we study the signature of the invisible decay of Higgs 
via the $ZH$ channel, where $Z$ can decay both leptonically and 
hadronically, $Z \to \ell\bar{\ell},  b \bar  b$.  
It is well known from an experimental point of view that the
leptonic channel is comparatively  cleaner than the hadronic
channel consisting of b-jets.  However we simulate both these channels
to find the detectability of an invisible Higgs decay.
In the following, we describe our simulation for both the final 
states.  
\\
(a) {\underline {Z $\to \ell\bar{\ell}$}}
\\
Here the final states consist of two leptons with opposite charge 
and same flavor and with a considerable 
amount of missing transverse momentum due to the Higgs decay into
invisible particles.  
    
The main dominant SM backgrounds are expected from the following
 processes,
\begin{enumerate}
 \item $ZZ$ production with one $Z$ decaying to neutrinos and the
 other $Z$ decaying leptonically.  Clearly, this background has exactly
identical  characteristics to the signal. 
 \item $WZ$ production followed by the leptonic decays of both the
 $W$ and $Z$,  giving rise to $\ell\nu_{\ell} \bar \ell \ell$ where one of
 the leptons is lost.
\item $WW$ production with both W bosons
 decaying leptonically,  $W \to \ell \nu_{\ell}$.   
\item Top pair production, $t\bar t \rightarrow W W b \bar
 b \rightarrow  l \nu_l \bar l \bar{\nu_l} b \bar b $ which may
 appear signal-like if the b-jets escape detection. 
\end{enumerate}
 The Higgs being heavier in comparison to the
 particles in the background processes other than the top quark,
 gives rise to a harder $\pt$. 
Therefore, by demanding a large $\pt$ one can efficiently reduce
 backgrounds.  In the signal topology, an added advantage 
is that the invariant mass of two leptons peaks
around the mass of the $Z$ boson.  Hence requiring the di-lepton
 invariant mass to be around
the mass of the $Z$ boson, it is possible to
 suppress backgrounds partially except for the $ZZ$ process.  
Since the $Z$ boson and the Higgs are more likely to be  produced
 back to back,  the transverse mass of the 
di-lepton system and the $\pt$, defined as,
\begin{equation}
 M_T^{l\bar{l}}=\sqrt{p_T^{ll}\pt\left(1-cos\phi(E_T^{ll},\pt) \right)},
\end{equation}
has a   softer distribution for all background processes.  
Therefore, demanding a large value for this variable 
enables us to eliminate backgrounds substantially.  

As before, we use 
{\tt MadGraph}\cite{Alwall:2011uj} to generate both the signal and
background processes which are
 subsequently passed through {\tt PYTHIA6} \cite{Sjostrand:2006za} for 
event generation including showering.  We apply the following
 set of cuts in our simulation for the event selection and as well as
  suppressing the background events. 
\begin{enumerate}
\item Select leptons with $p_T^{\ell}>10$~GeV and $|\eta_{l}|<3$.  
The isolation of lepton is ensured by looking at the total 
transverse energy $E_{T}^{ac}\le 20 \%$ of the $p_T$ of lepton,
where $E_{T}^{ac}$ is the scalar sum of the transverse energies of 
jets within a cone of size $\Delta R(l,j) \le 0.2 $ between the jet
 and the lepton.  

\item Since final states are hadronically quiet, 
vetoing events consisting jets, with $p_T> 30$~GeV and $|\eta|<4$
are useful in eliminating certain fraction of backgrounds. 
\item
 Azimuthal angle between two leptons, $\cos\phi_{\ell\bar{\ell}}>0$ 
and transverse mass between two leptons and $\pt$, $M_T^{l\bar{l}}>$
150 (200)~GeV for 8 (14) TeV energies.   
\item
Missing transverse momentum, $\pt>$100~GeV. 
\item
Di-lepton invariant mass, 
$|M_Z-m_{\ell\bar{\ell}}|< 10$~GeV.  
\end{enumerate}
For 14 TeV LHC energy, the strategy of simulation is not 
significantly different as no additional effects occur.  
The same set of cuts with similar thresholds are used with the 
only exception of $M_{T}^{\ell\bar{\ell}}$ where 200 GeV is used 
instead of 150 GeV.  In Table~\ref{dilepton}, we display
  cross sections for both signal and backgrounds for 8 and
 14 TeV  energies before and after cuts.
 In each column, numbers on left stand
 for the production cross sections corresponding to energies
 as shown in the respective  columns.  For both energies, we find
 that $M_T^{l\bar{l}}$ and $\pt$ play a very useful role in suppressing 
the backgrounds.  The kinematics of $ZZ$ process is identical to that
 of the signal process although there is a moderate mass 
difference (35 GeV) between the $Z$ and the Higgs boson, resulting in a 
similar effect of cuts on both signal and $ZZ$ background. As a 
consequence,  $ZZ$ turns out to be the dominant 
irreducible background. This channel was studied  extensively 
in an earlier study for 14 TeV LHC energy \cite{Godbole:2003it}.
 Here we have revisited the analysis for 14 TeV LHC energy
and performed an optimization of  cuts. 
 The numbers on the right hand side of 
each column show the final cross sections after being 
multiplied by acceptance efficiencies. For 8 TeV energy  with
 an  integrated luminosity of $\lumi$=20 $ \rm \invfb$ we find
 S/$\sqrt{B}\sim$ 3.5
 which implies  a hint of the invisible Higgs signal.  
However, for 14 TeV energy with $\lumi$=50 $\rm \invfb$ one can 
observe the invisible signal with signal significance of $\sim$8.  
Note that the estimations are  based on LO cross sections. 
However, we note that the K-factors for vector boson production and
$\rm t\bar{t}$ are 1.6-1.7~\cite{Campbell:2011bn,Kidonakis:2011tg} while for the
 signal process it is 
1.3\cite{Han1991167,PhysRevD.47.2730,PhysRevD.47.2722} respectively.
 Hence we do not expect any major 
changes in our results.
\begin{table}
\begin{center}
\begin{tabular}{|l|lr|lr| }
\hline
Process & \multicolumn{2}{c|}{8 TeV} & \multicolumn{2}{c|}{14 TeV} \\ 
  & Production & After Cuts & Production & After Cuts \\
  & C.S[pb]  &  C.S[fb]  & C.S[pb] & C.S[fb]   \\    
\hline
ZZ & 4.79 &6.7 & 10.1 & 17.6 \\
WZ & 12.6 & 1.8 & 47.3 & 3.8 \\
WW & 33.8 & 0.3 & 69.4 & 2.3 \\
$t\bar t$ & 115 & 0.1 & 480 & 0.95 \\
\hline
Total Bg & & 8.9 & & 24.7 \\
\hline
ZH & 0.3 & 2.3 & 0.64 & 5.6 \\
\hline
\end{tabular} 
\caption { Event summary for the dilepton+$\pt$ final states. 
 In the second and third columns, the  cross sections for
signal and backgrounds
before and after selection cuts, as described in the text, 
are presented for 8 TeV and 14 TeV center of mass energies respectively.}
\label{dilepton} 
\end{center}
\end{table} 

(b)\underline {Z $\to b \bar b$}   
\\
In this section we explore the possibility of detecting invisible
Higgs decay channel by identifying two b-jets arising from 
$Z$ boson decay.  We analyze this channel
following two methods.  In the first method,  b-jets are identified
 by using the standard jet clustering algorithm and in the second 
method, the jet substructure technique\cite{Butterworth:2008iy} is used
to reconstruct . 
 However, in both cases the dominant SM backgrounds arise from:
\begin{enumerate}
 \item  irreducible background from $ZZ$ production with one $Z$ 
decaying  to neutrinos and the other $Z$ decaying to b quarks. 
 \item The production of $Z$ boson in association with two b
 quarks  and the $Z$ boson decaying to neutrinos,
$(Z b \bar b \rightarrow \nu \bar{\nu} b \bar b)$.
\item $WZ$ production with the W decaying leptonically, and the $Z$ 
decaying to b-quarks and the lepton is lost, $(WZ\rightarrow l\nu_l b \bar b )$.
 \item $t \bar t$ production where two b-jet from top decays are 
identified and rest of the event objects are lost. 
 \item $W$ boson produced in association with $b$ quarks ($Wb\bar{b}$) where  
W decays leptonically and the lepton is not identified. 
\end{enumerate}
The event topology of this channel is not significantly different from the 
di-lepton final state as discussed above, and hence we 
apply similar type of cuts.  
Absence of any detectable hard lepton in the final state
leads us to apply a lepton veto to reduce  backgrounds, in 
particular from $t \bar t $, $WZ$ and $W b \bar b$ production.  
As before, $M_T^{b\bar{b}}$, the transverse mass between two b-jets
and $\pt$ distributions of the 
backgrounds are soft. Therefore, selection of signal events corresponding to
large values of these 
kinematic variables helps to remove significant fraction of the 
backgrounds. Moreover, we construct another useful variable, 
$R_T$, to remove large amount of the QCD.   
\cite{Guchait:2011fb,Chatterjee:2012qt,Ghosh:2012dh,Ghosh:2012mc,Byakti:2012qk}. 
This variable is defined as,
\begin{equation}
 R_T=\frac{p_{T_{b_{j_1}}}+p_{T_{b_{j_2}}}}{H_T},
\end{equation}
where $H_T$ is the scalar sum of the transverse momenta of all
detected jets including all non-tagged jets. 
 Since one expects less  non-tagged jet activity in the signal, $R_T$ would
 tend to have larger values
($ \sim$ 1) as compared to the events  arising from QCD and other
 backgrounds. Naturally, requiring $R_{T}$ to have a large value($\sim$ 1),
 leads to a substantial suppression of backgrounds, particularly for
 those due to QCD processes.

We simulate as before the signal and backgrounds using 
{\tt MadGraph} \cite{Alwall:2011uj} applying the
following set of selection  cuts:

\begin{enumerate}
\item 
Select b-jets by performing a matching between b quarks and jets
 using matching cone $\Delta R=0.3$ and finally multiply 
a b-tagging efficiency of 0.6 \cite{bjet} for each of the b-jets. 
In the jet substructure method we employ mass drop techniques 
described in \cite{Butterworth:2008iy} to find the subjets which
 are also identified as a b-like jets by flavor matching. 
\item 
Veto events with leptons, where $p_{T}^{l} > 10$~GeV and $|\eta_{l}|< 3$. 
\item 
Select dijet events with both jets b-like and  ensure that
 $|M_{b\bar b}-M_Z|<$ 30~GeV.  
\item 
$\pt >$ 70~GeV. 
\item
$M_T(b\bar{b},\pt)>$200~GeV. 
\item
$R_T>$ 0.9.  
\end{enumerate}
In Table~\ref{zbb8} we present the final results of the simulation for
both methods for 8 TeV energy.  The second column presents the total
production cross sections corresponding to each processes
and subsequent columns show cross sections after applying the above
set of cuts.  
However, in both cases, for an integrated luminosity of
 $\lumi=20~ \rm \invfb$ the best  we can achieve is
S/$\sqrt{B} \sim$ 2.
\begin{table}[t]
\begin{center}
\begin{tabular}{|c|c|c|c|}
\hline
Process & Production C.S[pb]  & After Cuts C.S [fb] & After cuts C.S[fb] \\
        & & b jet cluster & b jet substructure \\
\hline
$ZZ$
        &4.79       & 2.26 & 1.92  \\ 
$W Z$
        &12.6       & 0.38  & 0.36 \\ 
$\nu\bar \nu b \bar b $
        &16       &  3.1  & 1.33 \\ 
$t \bar t$
        &115        & 0.48  & 0.52 \\
W$ b \bar b$ 
        &50.5      & 0.54  & 0.16 \\
\hline
Background & & 6.76 & 4.29 \\
\hline
$Z H$   & 0.3       & 0.8 & 0.72 \\        
\hline
\end{tabular}
\caption{ Event summary for the final states with b jet pairs and 
$\pt$ for 8 TeV  energy.    The last two columns show the final
 cross sections after
 all cuts as described in the text. }
\label{zbb8}
\end{center}
\end{table}
In Table~\ref{zbb14} as before, we present results for 14 TeV
LHC energy. Here also we find that we can achieve a 
modest S/$\sqrt{B} \sim$ 4
with an integrated luminosity of $\lumi$=100 $ \rm \invfb$.
However, for a very high luminosity option, e.g $\lumi$=300 $\rm \invfb$, 
for a moderate value of signal
events, one can expect to observe an invisible BR of Higgs 
$\sim$75\% or more.  Note that because of low b-jet 
acceptance efficiency
and irreducible ZZ backgrounds this final state yields a 
marginal sensitivity. 
\begin{table}[h!]
\begin{center}
\begin{tabular}{|c|c|c|c|}
\hline
Process & Production C.S[pb]  & After Cuts C.S [fb] & After cuts C.S[fb] \\
        & & b jet cluster & b jet substructure \\
\hline
$ZZ$
        & 10   & 5.56 &  2.47 \\ 
$W Z$
        & 26.7   & 3.5  & 1.44  \\ 
$\nu\bar \nu b \bar b $
        & 47.3  &  12.9  & 3.04 \\ 
$t \bar t$
        & 476   & 3.92  & 0.16  \\
W$ b \bar b$ 
        & 112   & 4.2  &  1.08 \\
\hline
Background & & 30. &  8.19 \\
\hline
$Z H$   & 0.64  & 2.  & 1.1 \\        
\hline
\end{tabular}
\caption{ Same as Table \ref{zbb8}, but for 14 TeV LHC energy. }
\label{zbb14}
\end{center}
\end{table}
As we see jet substructure method does not give 
substantially better results because of the fact that the $Z$ boson 
is not sufficiently boosted.
Like the dilepton scenario as explained before, our results do not change 
significantly with the inclusion of higher order cross sections by using
appropriate K factors for Wb$\rm \bar{b}$, Zb$\rm\bar{b}$ \cite{Campbell:2000bg,
Campbell:2001ik},$\rm t\bar{t}$ \cite{Kidonakis:2011tg}
 and for the signal \cite{Han1991167,PhysRevD.47.2730,PhysRevD.47.2722}. 

\section{Summary}
Recent discovery of a Higgs like resonance by both the  experimental
groups: CMS and ATLAS
has now spurred a series of investigations to determine whether it is
`a' Higgs boson and if so
it is `the SM' Higgs boson. Assuming that it is `a'  Higgs boson the
current experimental information still does not rule out the
possibility of BSM physics. Many BSM models predict decay of the Higgs
in the
invisible channel along with the usual SM decay modes. Such invisible
decay modes, if confirmed or ruled out, will allow us to indirectly 
probe  BSM physics. In this note we revisit the possibility of
looking for a Higgs boson decaying invisibly, for two  production channels
of the Higgs : the vector boson fusion channel (VBF) as well as the
associated production of Higgs with $Z$ (ZH),  for two different LHC
energies: 8 and 14 TeV.  In the $ZH$ case, we also investigate the
possibility of using the $Z \rightarrow  b \bar b$ channel.
 In Table \ref{excl} we summarize the lower limits
of $\rm BR_{inv}$ for various 
channels and for different energy and
luminosity options.
We find that for the Z($\rm \to b\bar{b}$)H
channel we fail to set any
limits for 8 TeV with 20 $\rm fb^{-1}$
and 14 TeV with 30 $\rm fb^{-1}$
luminosity.
 We note
that in the VBF channel the sensitivity is more than $5 \sigma$
for both the energies : 8 and 14 TeV for large invisible branching
ratios ($> 0.8$) for integrated
luminosity of $20~{\rm fb}^{-1}$ and $ 30~{\rm fb}^{-1}$, whereas at
$14$ TeV with $300~{\rm fb}^{-1}$
one can reach an invisible branching ratio as low as $0.25$. In the ZH
channel with dileptonic decay of the Z, the sensitivity with the planned
luminosity of $20~{\rm fb}^{-1}$ is limited at $8$ TeV and rises to $8
\sigma$ at $14$ TeV with $50~{\rm fb}^{-1}$. With the $b \bar b$
final state, with
$20 {\rm fb}^{-1}$ we can only reach $S /\sqrt{B} \sim 2$ at 8 TeV energy, where as
with high luminosity
($\sim 300~{\rm fb}^{-1}$) and at 14 TeV energy  we can probe the invisible decay at $5 \sigma$
level, for an invisible branching ratio above $0.75$.
 As observations indicate that the determination
of an invisible branching fraction of the Higgs
 at the LHC is  difficult to achieve, specially for
small invisible branching ratios, an electron-positron linear collider with the 
associated production of the Higgs  along with a $Z$ boson can provide an 
extremely clean channel in this regard
~\cite{Schumacher:2003SS,Richard:2007ru}.  

\begin{table}[h!]
\begin{center}
\begin{tabular}{|c|c|c|c|}
\hline
Process & 8 TeV(20 $\rm fb^{-1}$)   & 14 TeV(30 $\rm fb^{-1}$) & 14 TeV(100 $\rm fb^{-1}$) \\
\hline
$VBF$
        & 0.34 & 0.32 & 0.17 \\ 
$ Z(\to l^{+}l^{-})H$
        & 0.58 & 0.32 & 0.18 \\ 
$ Z(\to b\bar{b})H$(substructure)
        & -- & -- & 0.5 \\ 
$ Z(\to b\bar{b})H$(b-jet cluster)
        & -- & -- & 0.55 \\
\hline
\end{tabular}
\caption{ The 95 $\%$ exclusion limits 
for $\rm BR_{inv}$
corresponding to various channels at 8 and 14 TeV
LHC energies and luminosities.}
\label{excl}
\end{center}
\end{table}

\section{Acknowledgements}
DG acknowledges support from ERC Ideas 
Starting Grant n.  
279972 ``NPFlavour".
DG also acknowledges the hospitality at the Centre for High Energy 
Physics, IISc where part
of this work was completed. 
RMG  wishes to acknowledge the Department of Science and Technology of India,
for financial support  under the J.C. Bose Fellowship scheme under grant no.
SR/S2/JCB-64/2007. 
MG acknowledges John Alwall for help
regarding matching in the context of MadEvent and V. Ravindran for 
useful discussion
related with Higgs cross sections.   
KM acknowledges CSIR fellowship. KM also acknowledges the Department of High
Energy Physics, Tata Institute of Fundamental Research for the 
hospitality where part of the work was done.

\bibliography{invhiggs.bib}{}
\bibliographystyle{utphys.bst}

\end{document}